# An accurate and revised version of optical character recognition-based speech synthesis using LabVIEW

Prateek Mehta[a1], Anasuya Patil[a*]

[a]Electrical Computer Engineering, Birla Institute of Technology, Mesra, Ranchi, India

**Abstract**:

Knowledge extraction just by listening to sounds is known as a distinctive property. Visually impaired people are dependent solely on Braille books & audio recordings provided by NGOs. Owing to many constraints in above two approaches blind people can't access the book of their choice. As the speech form is a more effective means of communication than text as blind and visually impaired persons can easily respond to sounds. This paper aims to develop an accurate, reliable, cost effective, and user-friendly optical character recognition (OCR) based speech synthesis system. The OCR based speech synthesis system has been developed using Laboratory virtual instruments engineering workbench (LabVIEW).

## I. INTRODUCTION

According to WHO, 285 million people are estimated to be visually impaired worldwide: 39 million are blind and 246 million have low vision. About 90% of the worlds visually impaired live in developing countries. Around 23,000 people are visually impaired in Oman in which 85% of the people are unable to get their basic needs fulfilled to lead a better and easier life. Products such as ABBYY FineReader, Readiris, etc. are available in the market for English language OCR Products such as NaturalReader, etc. serves the purpose of TTS system for English language in the form of software. But they are very costly, there prices range from $300 to $500, making them unaffordable for the visually impaired students for which presented work is dedicated.

Visually impaired people can't learn as we people normally do. The only way to learn for them is through Braille script which is a tactile writing system used by blind & visually impaired. An A4 size page of a normal textbook is equivalent to 4 pages in Braille script. Due to this fact Braille script books are large, heavy, take more time to read and are not easily available as compared to normal textbooks.
So, the proposed converter is intended to do the same task which will eliminate the cumbersome procedure of printing large size Braille script books & manual recording of normal textbooks [1], [2], [3], [4]. This converter will open new doors of Knowledge Sea for visually impaired as they can listen to any book of their choice [5].

## II. Related Work:

This work is related to existing research in text detection from general background or video image. S. K. Singla and R.K.Yadav [6] is OCR based speech synthesis system using LabVIEW. It contains two-part optical character recognition and text to speech conversion. The OCR software is developed with IMAQ Vision for LabVIEW software- developing tool and it uses a commercial digital scanner as image acquisition device. IMAQ Vision OCR software is a PC-based character recognition tool for use with IMAQ Vision for LabVIEW. For speech synthesis in LabVIEW the ACTIVE X sub pallet in Communication pallet and its functions to exchange data between applications. In this proposal only .wav files are supported from the beginning of OCR to TTS which results in excessive memory consumption and slow processing. Further this can only be used for specific fonts and the character size has to be above 48.

Shalin A. Chopra et al. [7] approach, The OCR system sends the ASCII representation of the character to the ARDUINO UNO board which correspondingly actuates the Braille display. When the visually impaired person touches the Braille display, they could understand the printed text. This can only read the text up to the font size of 28 and can only project one character at a time. It takes a total delay of 30 milliseconds between each projection which can be extremely slow for large amount to text.

Recently, Joshi et al. [8] an electronic pen aiding visually impaired to read the text is proposed. They suggest for the use of a pen which consists of a camera, conversion software, word repository, text-to-audio converter. But the text to audio conversion requires huge database. The system has to recognize each character and thus the word then find the word's appropriate audio file from the huge database and then transmit them.

Research Work done by [1][9][10][11] and a comprehensive review [12] talks about the automation that can be brought to ease out the text detection and optical recognition but fails to implement the necessary accuracy in detection, the speed at which it should be done and is also limited the font size that can be used to make it work.

### III. CONSTRUCTION OF OCR SYSTEM

*A. Pre Processing*

The file is taken in PDF format and is converted into the required text.

This is done by parsing PDF using PDFBox. It is another Java PDF library. The PDFBox library is an open-source Java tool for working with PDF documents. This tool allows manipulation of existing documents and the ability to extract content from documents.

*B. Character Extraction*

After the text form is extracted form PDF format, desired color can be set in which the text will be displayed.

*C. Recognition*

The extracted text further goes for search string option. Here the required text can be typed in the given box and if the string is matched, the matched word is highlighted by a given color. If the required word is not found, no action takes place.

*D. Post Processing*

The achieved OCR can be saved as a text file for the speech processing or further reference.

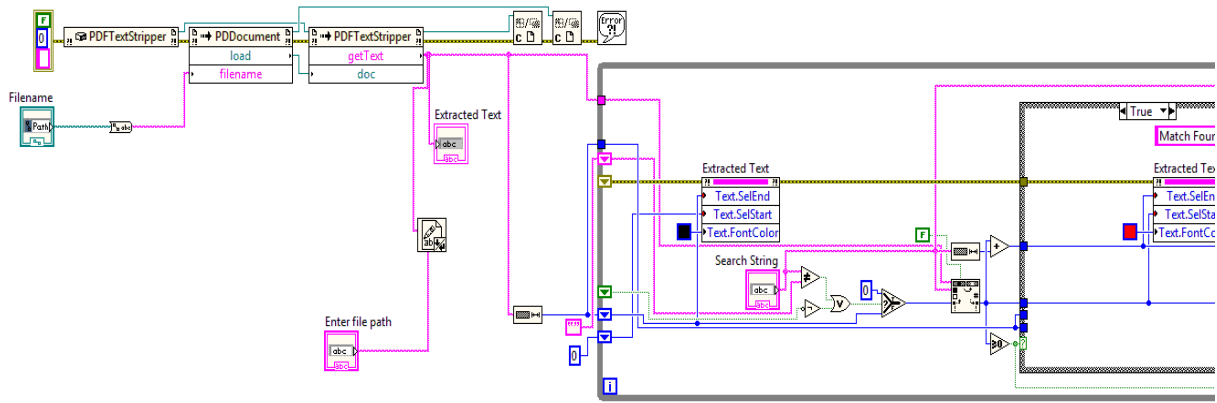

Fig 1. Construction of OCR system

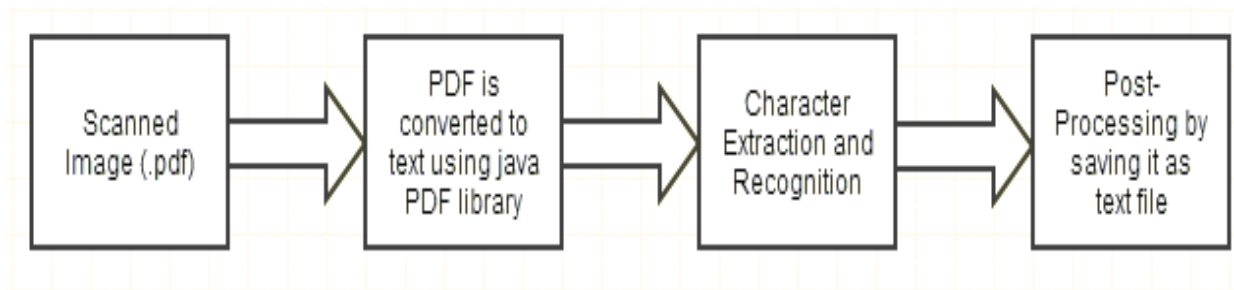

Fig 2. OCR Block Diagram

## IV. CONSTRUCTION OF TTS SYSTEM

In text to speech module text recognized by OCR system will be the inputs of speech synthesis system which is to be converted into speech.

Here for speech system, Microsoft Speech SDK 5.1 is used. The LLB VIs call Microsoft Speech SDK 5.1 and retrieve the voice and audio output information which is available for the computer. This VI allows you to select the voice and audio device you would like to use, enter the text to be read, and adjust the rate and volume of the selected voice.

The VI created offers two options for speech system, we can either extract the OCR text file or manually type the content to be converted into speech.

Along with the ability to adjust the speech rate and the volume, we can also the pause the audio at the desired stage and can start it from that point at any period of time.

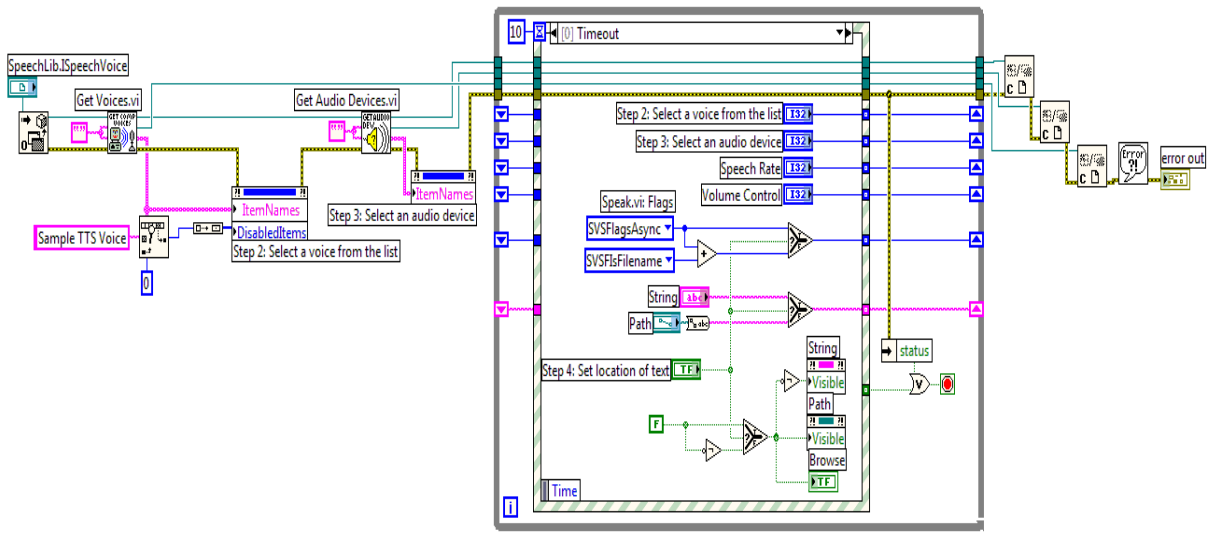

Fig 3. Construction of TTS system

**The NI LabVIEW™ Program**

Using LabVIEW, and injecting the program into the NI-myRio, allowed the connection between the hardware device and the computer software. The program initiates the Wi-Fi connection. It consists of several blocks, and each has its own functionality. NI-myRio firsts listen for any file stored by a scanner then ensures a TCP/IP connection specifying the IP address and the port at which the connection would be established. Then an interrupt line occurs to call-back VI from the main LabVIEW software to write the data received from scanning system that is able to acknowledge the movements done of the data and forwards the information to the application in order to update the file to be converted. OCR reads the path, gathers data from the PDF, converts it into text and sends this data to the TTS block to convert it into audio form. TTS configures every character to get distinct words in the sentences and through the help of the inbuilt system language, it gives out the speech signal waves reading out the entire given text. The whole program is enclosed in a while loop that ensures continuous flow of data with an appropriate delay after every sentence. This is shown in Figures 4 and 5, the blocks that are used for image processing to the speech conversion.

# V. RESULTS AND DISCUSSION

Experiments have been performed to test the proposed system developed using LabVIEW 13.0 version. The developed OCR based speech synthesis system has two steps:
a. Optical Character Recognition
b. Speech Synthesis

## A. Optical Character Recognition

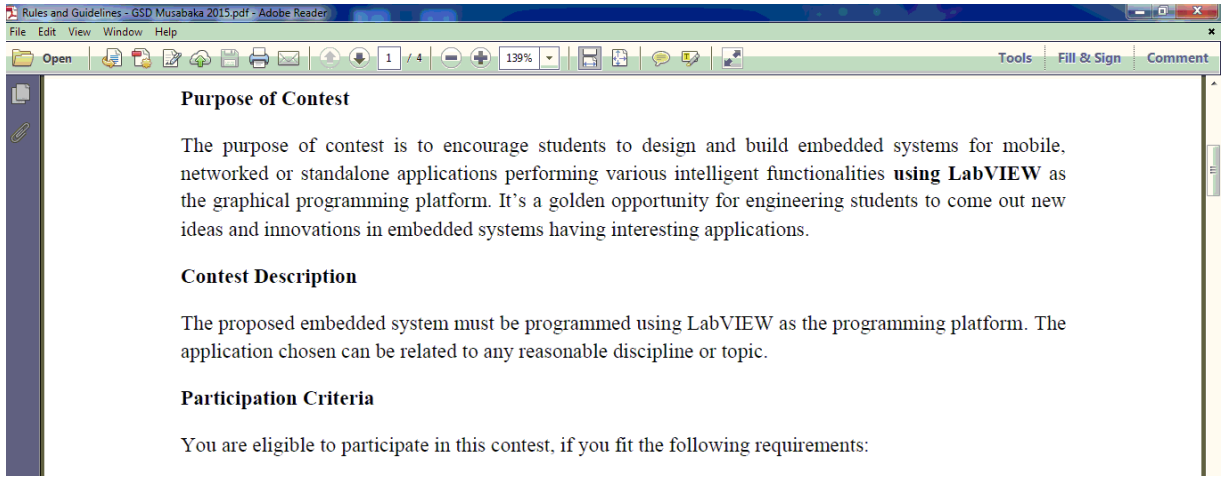

Fig 4. PDF file example

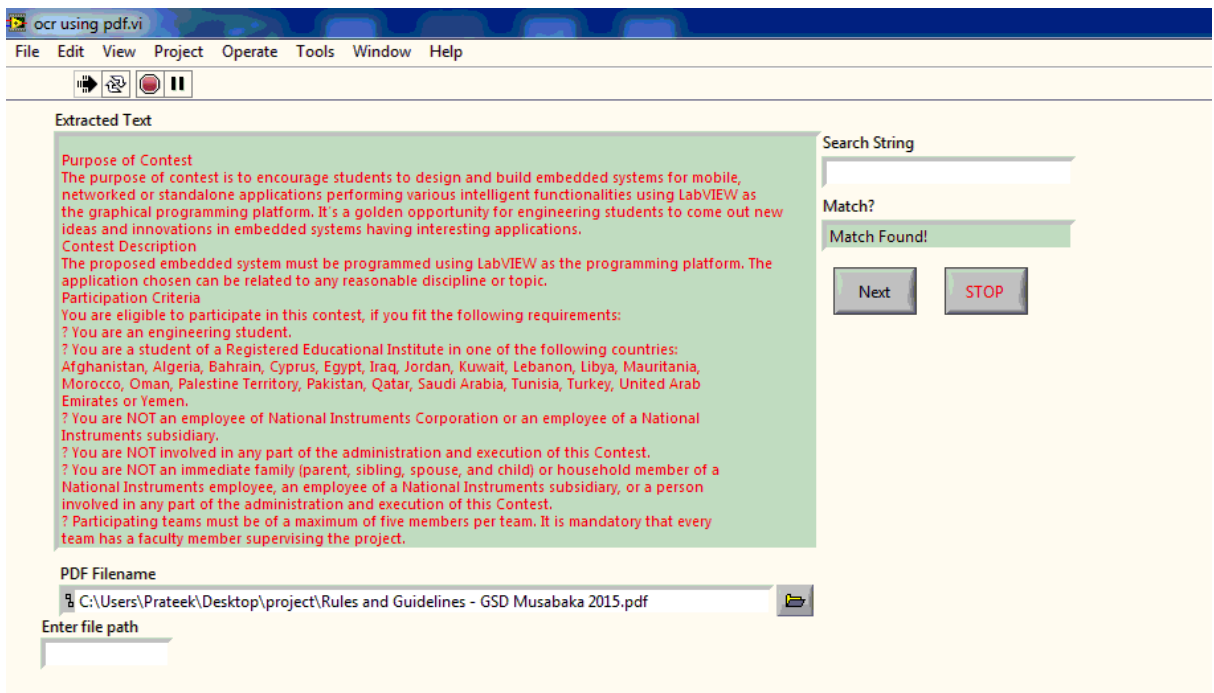

Fig 5. Result from OCR system

### B. Speech Synthesis

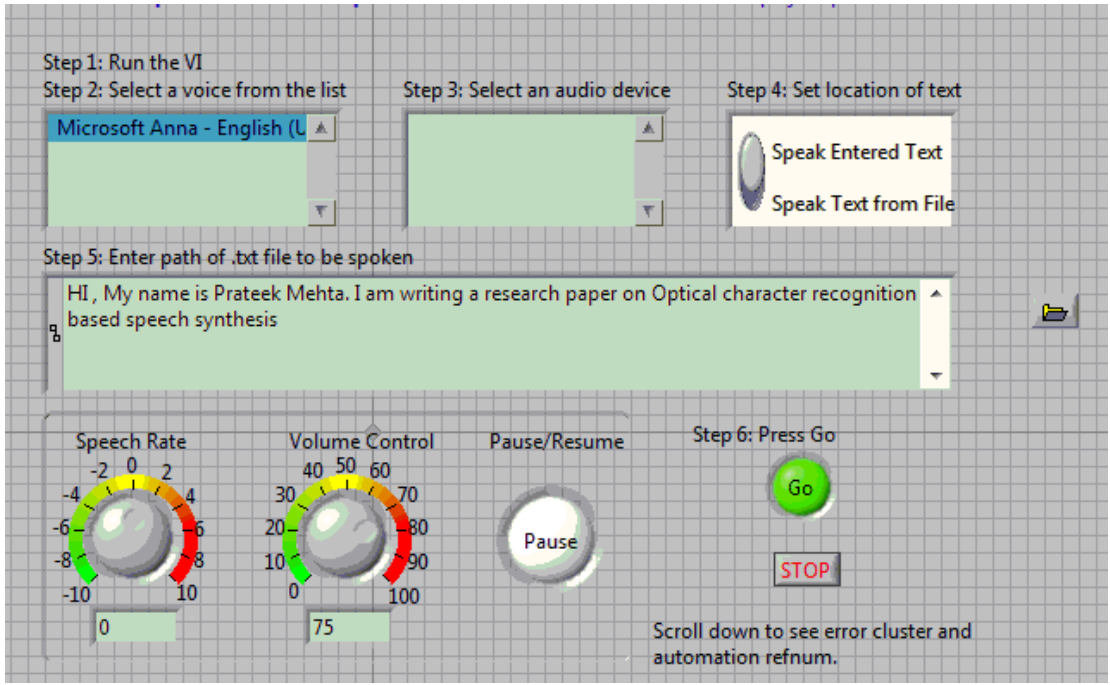

Fig 6. Result of TTS system with Entered Text

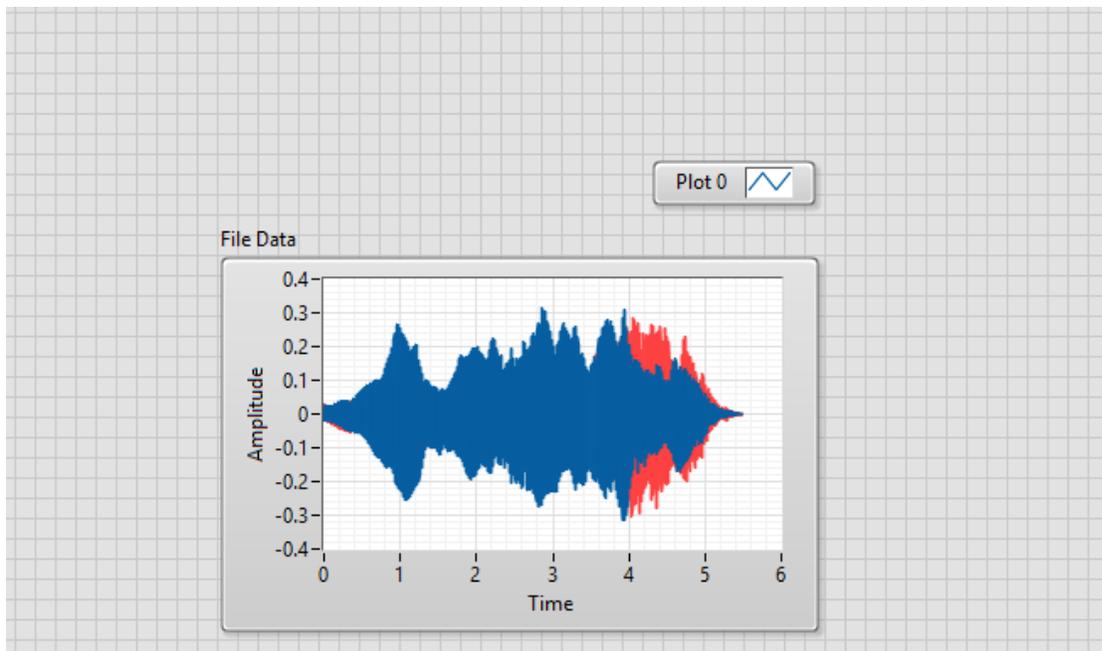

Fig 7. Sound Waveform

## VI. CONCLUSION

In this paper, an OCR based speech synthesis has been successfully implemented for English on LabVIEW 13.0 platform. Therefore, presented work can be implemented in blind schools and colleges using readily available hardware such as personal computer and scanner (.txt & .pdf format) along with readily available books format (such as text or .pdf). Also, the existing problem discussed above is solved successfully which is almost 100% accurate. The developed system consists of OCR and speech synthesis. The developed OCR based speech synthesis system is accurate, reliable, user friendly, cost effective and gives the result in the real time.

Moreover, the program has the required flexibility to be modified easily if required.

The reason why NI products were chosen is mainly the time savings and the simplicity they bring in. The NI-myRio operates real time LabVIEW OS which makes it very fast and reliable. The NI LabVIEW graphical programming language makes faster and more intuitive programming. In addition, the built-in made the work on the project much easier without the hassle of other programming languages. The programming time was radically reduced, and the authors needed the spared time to focus on the hardware implementation.

Furthermore, the project is innovative as it brings new concept to the disabled, making people of different ages interested in enrolling into a better and developing world.

Following is the comparison with other similar projects done based on accuracy and the font compatibility for conversion.

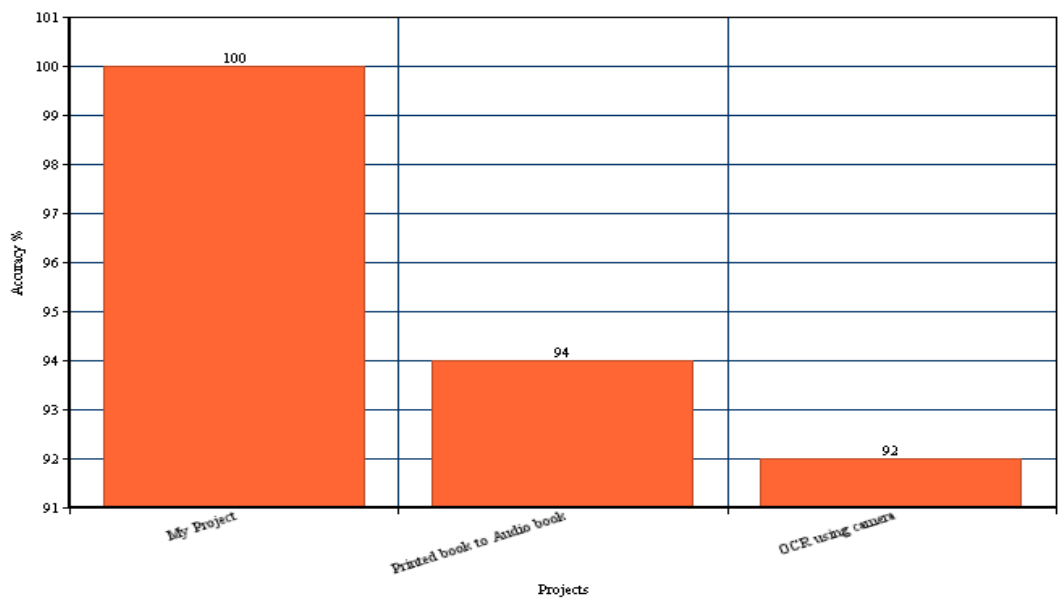

Fig 8. Comparison of Accuracy

Fig 9. Font Size Compatibility for conversion

## VII. REFERENCES

[1]  J. Yang, "An automatic sign recognition and translation system", doi: 10.1145/971478.971490.
[2]  J. Bliss, "A Relatively High-Resolution Reading Aid for the Blind," *IEEE Trans. Man Mach. Syst.*, vol. 10, no. 1, pp. 1–9, Mar. 1969, doi: 10.1109/TMMS.1969.299874.
[3]  P. Mehta *et al.*, "AI Enabled Ensemble Deep Learning Method for Automated Sensing and Quantification of DNA Damage in Comet Assay," *ECS Sens. Plus*, vol. 2, no. 1, p. 011401, Mar. 2023, doi: 10.1149/2754-2726/acb2da.
[4]  S. Namuduri *et al.*, "Automated Quantification of DNA Damage Using Deep Learning and Use of Synthetic Data Generated from Basic Geometric Shapes," *ECS Sens. Plus*, vol. 3, no. 1, p. 012401, Mar. 2024, doi: 10.1149/2754-2726/ad21ea.
[5]  J. Tuckova and M. Sramka, "ANN application in emotional speech analysis," *Int. J. Data Anal. Tech. Strateg.*, vol. 4, no. 3, p. 256, 2012, doi: 10.1504/IJDATS.2012.047819.
[6]  S. K. Singla and R. K. Yadav, "Optical Character Recognition Based Speech Synthesis System Using LabVIEW," *J. Appl. Res. Technol.*, vol. 12, no. 5, pp. 919–926, Oct. 2014, doi: 10.1016/S1665-6423(14)70598-X.
[7]  S. Chopra, "Optical Character Recognition." [Online]. Available: https://ijarcce.com/wp-content/uploads/2012/03/IJARCCE2G_a_shalin_chopra_Optical.pdf
[8]  J. K. A. V., V. A., M. R. S., M. P. T., and K. V. K. G., "PENPAL - Electronic Pen Aiding Visually Impaired in Reading and Visualizing Textual Contents," in *2011 IEEE International Conference on Technology for Education*, Chennai, India: IEEE, Jul. 2011, pp. 171–176. doi: 10.1109/T4E.2011.34.
[9]  Huiping Li, D. Doermann, and O. Kia, "Automatic text detection and tracking in digital video," *IEEE Trans. Image Process.*, vol. 9, no. 1, pp. 147–156, Jan. 2000, doi: 10.1109/83.817607.

[10] T. Sato, T. Kanade, E. K. Hughes, M. A. Smith, and S. Satoh, "Video OCR: indexing digital news libraries by recognition of superimposed captions," *Multimed. Syst.*, vol. 7, no. 5, pp. 385–395, Sep. 1999, doi: 10.1007/s005300050140.

[11] T. Dutoit, "High quality text-to-speech synthesis: a comparison of four candidate algorithms," in *Proceedings of ICASSP '94. IEEE International Conference on Acoustics, Speech and Signal Processing*, Adelaide, SA, Australia: IEEE, 1994, p. I/565-I/568. doi: 10.1109/ICASSP.1994.389231.

[12] P. Mehta, M. A. Mujawar, S. Lafrance, S. Bernadin, D. Ewing, and S. Bhansali, "Editors' Choice—Review—Sensor-Based and Computational Methods for Error Detection and Correction in 3D Printing," *ECS Sens. Plus*, vol. 3, no. 3, p. 030602, Sep. 2024, doi: 10.1149/2754-2726/ad7a88.